\RequirePackage{fix-cm}
\documentclass{svjour3}                     
\smartqed  
\usepackage{graphicx}
\usepackage{bm}
\usepackage{amsmath}
\usepackage{amssymb}
\usepackage{mathptmx}      
\usepackage{latexsym}
\usepackage{verbatim}
\usepackage{color}
\usepackage[english]{babel}

\usepackage{cite}

\journalname{Journal of Computational Electronics}

\begin{document}

\title{Comparison of electron and phonon transport in disordered semiconductor carbon nanotubes}

\author{H.~Sevin\c{c}li, T.~Lehmann,  D.\,A.~Ryndyk, and G.~Cuniberti}

\institute{H. Sevin\c{c}li$^{1,3}$, T.~Lehmann$^{1}$, D.\,A. Ryndyk$^{1,2}$, and G. Cuniberti$^{1,2}$ \at
$^{1}$ Institute for Materials Science and Max Bergmann Center of Biomaterials,
Dresden University of Technology,
01062 Dresden, Germany.\at
$^{2}$ Center for Advancing Electronics Dresden, TU Dresden, 01062 Dresden, Germany.\at
$^{3}$ Department of Micro- and Nanotechnology (DTU Nanotech), Technical University of Denmark, DK-2800 Kgs. Lyngby, Denmark
}

\date{Received: date / Accepted: date}

\maketitle

\begin{abstract} Charge and thermal conductivities are the most important parameters of carbon nanomaterials as candidates for future electronics. In this paper we address the effects of Anderson type disorder in long semiconductor carbon nanotubes (CNTs) to electron charge conductivity and lattice thermal conductivity using the atomistic Green function approach. The electron and phonon transmissions are analyzed as a function of the length of the disordered nanostructures. The thermal conductance as a function of temperature is calculated for different lengths. Analysis of the transmission probabilities as a function of length of the disordered device shows that both electrons and  phonons with different energies display different transport regimes, i.e. quasi-ballistic, diffusive and localization regimes coexist. In the light of the results we discuss heating of the semiconductor device in electronic applications. 


\end{abstract}

\section{Introduction}

Electron transport in carbon nanostructures, namely nanotubes and graphene nanoribbons, is in the focus of experimental and theoretical research during last years. In particular, the Anderson disorder model and localization of electrons in CNTs have been studied extensively in the past decade \cite{Latil04prl,Roche05prb,Charlier07rmp,Lherbier08prl}. It is important to note that also {\em phonon} thermal transport in low-dimensional systems is of central importance for applications. By properly controlling thermal properties it is possible to enhance the device performance. In electronic applications, high values of both electronic and phononic conductances are desired. On the other hand a low phonon conductance is required in order to have efficient thermoelectric energy conversion. In this paper we focus on the phonon thermal conductivity in disordered systems and compare the effects of disorder on the phonon conductivity with the effect on the electron transport. 

In semiconductor materials heat transport is governed mostly by phonons and it is previously shown that the thermal conduction is strongly influenced by device dimensions \cite{Dresselhaus08jvacscitechb}. For example, it was observed recently that phonon conduction can be reduced without effecting the electronic transport  importantly in Si nanowires which leads to an enhancement in the thermoelectric figure of merit by two orders of magnitude compared to its bulk value \cite{Hochbaum08nature,Boukai08nature,Markussen09prb}.
This decrease in thermal conduction is mainly due to the scattering of phonons at disordered surface whereas electron transfer is maintained by bulk-like states.
Thus, surface to volume ratio is a parameter to control the transport properties of these devices.
Graphitic allotropes have the exceptional property that they are one atom tick.
There is a growing interest in the field of phononic energy transport through carbon based materials.
It is shown that thermal conductance of nanotubes can be tuned by sliding the inner shell inside the outer shell \cite{Chang07apl}. 
Disorder induced localization is believed to be a possible explanation of the exponential dependence of thermal resistance on the telescoping distance.
Conductance is independent of the device length if the device is pristine.
Real systems, on the other hand, always include disorder which we model with a distribution in force constants.
Recently the effect of isotopic disorder on thermal conduction through nanotubes is studied theoretically \cite{Savic08prl,Stewart09nanolett} and good agreement with experiment \cite{Chang06prb} is achieved.

In this paper, we analyze the effects of Anderson disorder in semiconductor CNTs on electron and phonon transport. First we summarize the atomistic Green function method in calculating the transport properties. Then we investigate the dependence on length of the device and on the operating temperature. The energy dependence of transport regimes are shown to coexist and their role on device performance depending on system parameters is discussed.

\section{Model and Method}

\begin{figure}[b]
	\begin{center}
	\includegraphics[width=1.\textwidth]{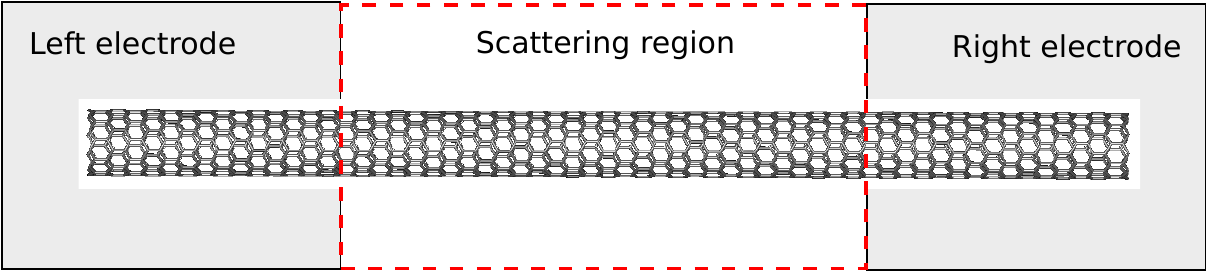} 
	\caption{Considered system: a carbon nanotube between electrodes.}
	\label{fig:cnt-fet}
	\end{center}
\end{figure}

As it is common in transport calculations, we apply the partitioning scheme and divide the system into left electrode, center (scattering region) and right electrode subsystems ($\alpha=L$, $C$, and $R$ respectively). Left and right electrodes are considered to be equilibrium, but with different electrical potentials $\varphi_L$ and $\varphi_R$, as well as different temperatures $T_L$ and $T_R$ (Fig.\,\ref{fig:cnt-fet}). 

\subsection{Electron transport}

\begin{figure}[t]
\begin{center}
\includegraphics[width=120mm]{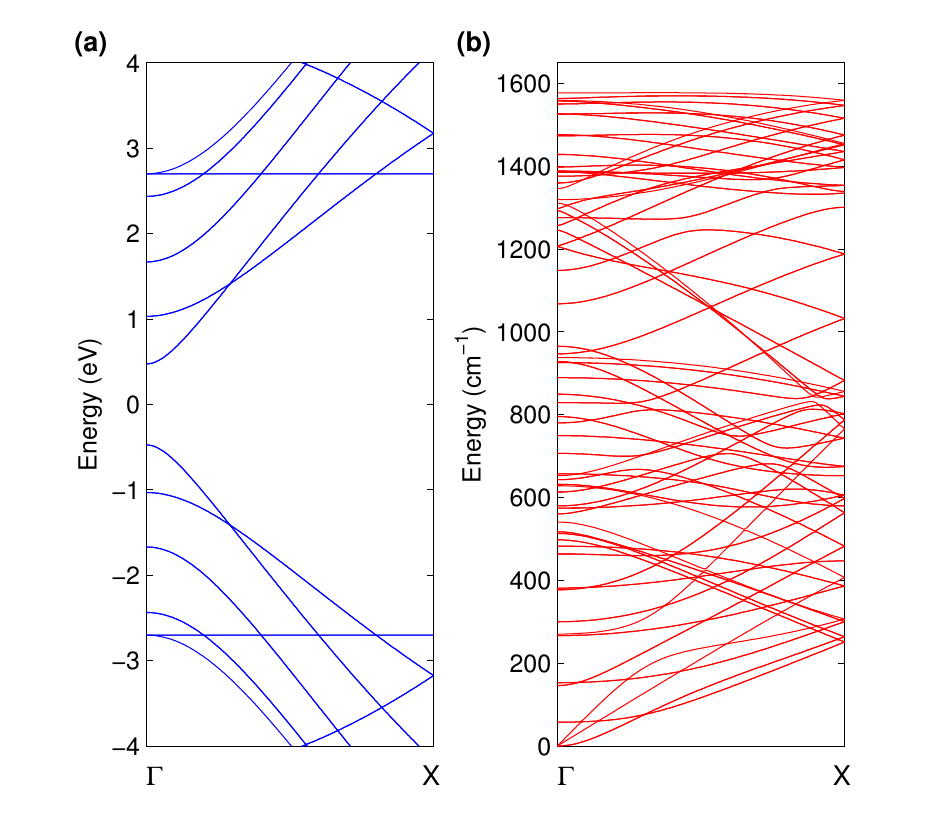}
\caption{Electron (a) and phonon (b) dispersions in pristine CNT(10,0).}
\label{fig:dispersion}
\end{center}
\end{figure}

We describe electron states in carbon nanotubes in the framework of the \mbox{$\pi$-electron} tight-binding model~\cite{Wallace47prb}. This approach is complementary to an ab initio one and allows to obtain physically transparent results. We write the first nearest neighbor single orbital tight-binding Hamiltonian of our system (for fixed spin $\sigma=\uparrow$ or $\sigma=\downarrow$, the spin index is omitted) as
\begin{equation}
   \hat H=\sum_{i} \epsilon_i\,c^{\dag}_{i}c_{i} +\sum_{ij} t_{ij}\,c^{\dag}_{i}c_{j},
\end{equation}
where hopping matrix elements $t_{ij}$ between lattice sites are given by $t_{ij}=t$ for nearest neighbor sites and the on-site energies $\epsilon_i$ can be chosen zero for pristine structures. We take below $t=2.7$ eV. As an example, the electron dispersion in ideal infinite CNT(10,0) is shown in Fig.~\ref{fig:dispersion}a.

In calculating the transport properties, we follow an atomistic approach and employ nonequilibrium Green functions within the Landauer formalism, and a decimation technique \cite{LopezSancho84jpf,LopezSancho85jpf} to obtain transmission functions ${\cal T}_{ph}(\omega)$ and ${\cal T}_{el}(E)$ for phonons and electrons, respectively. We refer the reader to Refs. \cite{Datta95book,Cuniberti05book,Ryndyk09inbook,Cuevas10book} for details of the Green function technique. We neglect the electron-phonon coupling in this work since it is shown that electron-phonon mean free path in nanoscale carbon tubes and ribbons is tens of $\mu$m even at room temperature~\cite{Gunlycke07prb}.

The electron current is given by the Landauer formula
\begin{equation}
  I=\frac{e}{h}\int_{-\infty}^\infty {\cal T}_{el}(E)\left[f_L(E)-f_R(E)\right]dE,
\end{equation}
where $f_{L(R)}$ is the electron distribution function in the left (right) electrode and the transmission function ${\cal T}_{el}$ is defined as
\begin{equation}
  {\cal T}_{el}(E)={\rm Tr}\left(\hat\Gamma_L(E)\hat G^R_C(E)\hat\Gamma_R(E)\hat G^A_C(E)\right),
\end{equation}
where the Green matrix functions in the central region $\hat G^{R(A)}_C$ and level-width functions $\hat\Gamma_{R(L)}$ are determined by the Hamiltonian of the central region ${\hat H}_C $ and the contact self-energies ${\hat\Sigma}_s$
\begin{equation}
  {\hat G}^R_{C}(\epsilon)=\left[(\epsilon+\mathrm{i}0^+){\hat I}-{\hat H}_C-{\hat\Sigma}_L-{\hat\Sigma}_R\right]^{-1},
\end{equation}
\begin{equation}
  \hat\Gamma_{s=L,R}=i\left(\hat\Sigma^R_s-\hat\Sigma^A_s\right)=-2{\rm Im}\hat\Sigma^R_s.
\end{equation}
While we use the tight-binding model, all Green functions are matrices in the atomistic basis.

\subsection{Phonon transport}

In the harmonic approximation the vibrational Hamiltonian can be written as
\begin{equation}
 	H_{ph}=
	\sum\limits_\alpha
	H_\alpha
	+
	\left(
	u^L | \hat K^{LC} | u^C
	\right)
	+
	\left(
	u^R | \hat K^{RC} | u^C
	\right),
\end{equation}
where in the first term
\begin{equation}
H_\alpha=
	\frac{1}{2}\left(\dot{u}^\alpha|\dot{u}^\alpha\right)
	+
	\frac{1}{2}\left(u^\alpha | \hat K^{\alpha\alpha} | u^\alpha\right)
\end{equation}
is the Hamiltonian of the decoupled subsystem $\alpha$,
the second and third terms stand for the coupling between the central region and the reservoirs.
Here $\left.|u^\alpha\right)$ is the vector, and $\left(u^\alpha|\right.$ is its transpose, of mass renormalized displacement coordinates $u_i^\alpha=\sqrt{m_i}x_i^\alpha$, $x_i^\alpha$ being the $i^{\mathrm{th}}$ degree of freedom of subsystem $\alpha$ with $m_i$ the mass associated to this degree of freedom. 
$K_{ij}^{\alpha\beta}$ is the matrix element representing the coupling between mass renormalized coordinate $i$  of subsystem $\alpha$ with $j$ subsystem $\beta$, and
$K_{ij}^{\alpha\beta}=k_{ij}^{\alpha\beta}/\sqrt{m_im_j}$ with $k_{ij}^{\alpha\beta}$ being the spring constant in direct coordinates.
In our calculations, we use the fourth nearest neighbor force constant approximation (4NNFC), which yields phonon dispersions in agreement with density functional theory (DFT) calculations for graphene and carbon nanotubes \cite{Saito88book,Zimmermann08prb,Sevincli0905.3815}.
For the case of GNRs, modification of force constants for the edge carbon atoms will improve the results, in the sense that it will result in a blue shift in the phonon density of states. Theoretical calculations show that hydrogenated and dehydrogenated ribbons have very similar phonon dispersions~\cite{Yamada08prb}. Therefore, we neglect the effects of hydrogenization in our phonon calculations.
A subsequent reduction of lattice thermal conductance and a further but minor improvement to our results in reducing lattice thermal conductivity can be expected, but it is neglected for the sake of simplicity. As an example, the phonon dispersion in ideal infinite CNT(10,0) is shown in Fig.~\ref{fig:dispersion}b.

Phonon heat transport in mesoscopic systems at low enough temperatures (lower than the Debye temperature $T_D$, which is about 2300 K in graphene and is of the order of 1000K in CNTs) is essentially quantum and at $T\ll T_D$ the heat flow is determined by coherent transport of noninteracting phonons. In this case the Landauer approach can be used. The thermal conductance in the limit of small temperature difference $dT=T_L-T_R$ is defined as
\begin{equation}
	\kappa_{ph}(T)=
	\int\limits_0^\infty
	\frac{d\omega}{2\pi}
	\,\hbar\omega
	\frac{\partial f_B(\omega,T)}{\partial T}
	{\cal{T}}_{ph}(\omega),
	\label{eqn:conductace}
\end{equation}
where ${\cal T}_{ph}(\omega)$ is the phonon transmission function, $f_B$ is the Bose distribution with $\omega$ being the phonon frequency and $T$ being the temperature~\cite{Angelescu98sm,Rego98prl}.

To actually calculate the phonon transmission function for nonideal nanotubes we use the atomistic single-particle Green function method~\cite{Ozpineci01prb,Mingo03prb,Segal03jcp,Mingo06prb,Wang06prb,Yamamoto06prb,Galperin07prb2,Wang08epjb,Dubi11rmp,Nikolic12jcel}, equivalent to the Meir-Wingreen method for electron transport~\cite{Meir92prl}.

${\cal{T}}_{ph}(\omega)$ is determined by the matrix retarded phonon Green function in the central region $\hat D_C(\omega)$ and the phonon level-width functions $\hat\Gamma^{ph}_{L(R)}(\omega)$ of the left and right electrodes:
\begin{equation}
{\cal{T}}_{ph}(\omega)=
\mathrm{Tr}
\left[\hat\Gamma^{ph}_{L}(\omega)\hat D_C^\dag(\omega)\hat\Gamma^{ph}_{R}(\omega)\hat D_C(\omega)\right].
\end{equation}
The retarded Green functions for subsystems in the absence of coupling between the central system with the electrodes are defined as
\begin{equation}\label{D0}
D_\alpha^{(0)}(\omega)= \left[(\omega+\mathrm{i}0^+)^2\hat I - \hat K^{\alpha\alpha} \right]^{-1}
\end{equation}
with $\hat I$ being the identity matrix.  The matrix inversion (\ref{D0}) can not be applied directly for semi-infinite electrodes and the surface Green functions are calculated numerically using the iterative method \cite{LopezSancho84jpf,LopezSancho85jpf}. 

The retarded phonon Green function for the central region in the presence of electrodes reads
\begin{equation}
	\hat D_C(\omega)= 
	\left[
	(\omega+\mathrm{i}0^+)^2\hat I -\hat K^{CC} - \hat\Pi_L(\omega) - \hat\Pi_R(\omega)
	\right]^{-1},
\end{equation}
where the polarization operator (phonon self energy) due to coupling to the reservoir $\alpha$ is
\begin{equation}
\hat\Pi_\alpha(\omega)=\hat K^{C\alpha}D_\alpha^{(0)}(\omega)\hat K^{\alpha C}.
\end{equation}

Finally, the level-width function is determined by the relation
\begin{equation}
\hat\Gamma^{ph}_\alpha(\omega)=\mathrm{i}\left(\hat\Pi_\alpha(\omega)-\hat\Pi^\dag_\alpha(\omega)\right).
\end{equation}

\section{Results for disordered nanotubes}

\begin{figure}[b]
	\begin{center}
	\includegraphics[width=0.71\textwidth]{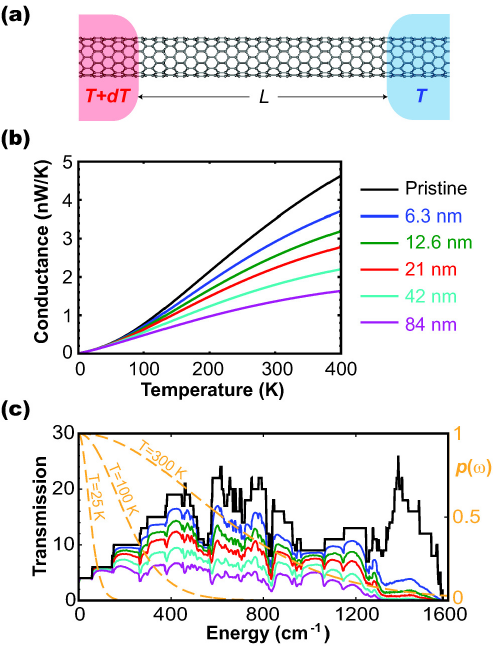} 
	\caption{
	(a) Schematics of the system under consideration.
	(10,0) CNT is placed between reservoirs which have a temperature difference $dT$.
	We include Anderson type disorder at the central region whose length $L$ is varied.
	(b) Thermal conductance as a fnction of temperature $T$ for (10,0) CNTs with diffrent lengths.
	(c) Transmission versus energy for CNTs with diffrent lengths $L$ (solid curves).
	Black curve is the transmission function of pristine (10,0) CNT.
	Dashed curves are the prefactors $p(\omega,T)$ for $T=$25, 100, 300 K (see text).
	(Same color code is used in subplots (b) and (c)).
	}
	\label{fig:trans}
	\end{center}
\end{figure}

\begin{figure}[t]
	\begin{center}
	\includegraphics[width=0.8\textwidth]{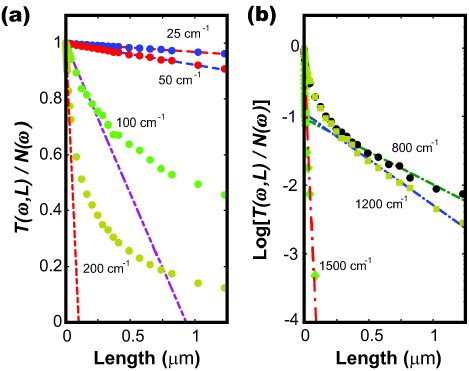}
	\caption{
	Different regimes of phonon transport through (10,0) CNTs.
	Normalized transmission values are plotted as a function of length for various energies.
	At low energies (25 cm$^{-1}$, 50 cm$^{-1}$) ${\cal{T}}(\omega)/N(\omega)$ can be fit to a line (a).
	Intermediate energies like 100 cm$^{-1}$ and 200 cm$^{-1}$ deviate from the linear behavior and show $1/L$ behavior (a).
	High energy phonons are best described with an exponential law (b).
	}
	\label{fig:regime}
	\end{center}
\end{figure}

Anderson disorder represents random small-scale fluctuations of system parameters. In this paper we consider simple local and uncorrelated disorder. For phonons we introduce Anderson type disorder by a random distribution of $m_i$ at the $C$ region, in the range plus/minus 10\% of the original mass, which transforms to a distribution of disordered coupling matrix elements $K_{ij}^{CC}$. Green functions of the disordered region are obtained using decimation techniques and the transmission functions are obtained by averaging over 250 sample configurations. The results are presented in Fig.\,\ref{fig:trans} and Fig.\,\ref{fig:regime}. 

First of all we note that, unlike its electronic counterpart, phononic heat transmission takes place not only in a small energy window but phonons of all energy values contribute to the conductance.
In this sense, low energy phonons play a special role in phonon thermal conduction in both pristine and disordered systems as it will be discussed below.
We consider (10,0) CNT which is placed between reservoirs made of the same material [Fig.~\ref{fig:trans}(a)].
The reservoirs are kept at different temperatures $T$ and $T+dT$.
We introduce Anderson type disorder in the central region and analyze the dependence of the transmission function and thermal conductance to disorder for CNTs with different lengths.
The transmission function for the pristine CNT reflects the fact that each phonon branch is contributing to transport with unit probability.

Namely, the transmission function at energy $\omega$ is the number of phonon branches $N(\omega)$ crossing at this energy  each acting as an independent transmission channel [Fig.~\ref{fig:dispersion}(b)]. 
Increasing the length of the $C$-region, the transmission is reduced but the low energy transmission is almost unaffected by disorder [Fig.~\ref{fig:trans}(c)].
On the other hand high energy phonons are extremely sensitive to disorder in the sense that the  transmission is blocked to a great extend even for the shortest regions with disorder.
We also note that disorder induced scattering is more important near van Hove singularities in the density of states.
We observe that thermal conductance drops as the length of the central region is increased but this drop is significant only at high temperatures [Fig.~\ref{fig:trans}(b)].
As the temperature increases the difference in conductance for different lengths increase and then saturates.
The low temperature thermal conductance is insensitive to disorder because of two reasons.
First, because the transmission of low energy phonons itself is insensitive to disorder.
Second, the term $p(\omega,T)=\omega\,\partial f_B(\omega,T)/\partial T$ in the integrand of Eq.\,(\ref{eqn:conductace}) is filtering out the contribution of high energy phonons to low temperature conductance.
At high temperatures $p(\omega)$ flattens and the filter effect diminishes.

In order to identify the different transport regimes, we analyze ${\cal{T}}_{ph}(\omega,L)/N(\omega)$ for a number of $\omega$ values, and as a function of $L$.
Transmission amplitudes are normalized in this way using their pristine values $N(\omega$) in order to enable comparison.
In Fig.\ref{fig:regime}(a), it is shown that low energy transmission ($\omega=$25, 50 cm$^{-1}$) decreases linearly with $L$ reflecting the fact that low energy phonons display a quasi-ballistic behavior.
For higher energies ($\omega=$100, 200 cm$^{-1}$) the transmission function deviates from the linear fit to short device transmission (see Fig.~\ref{fig:regime}(a)) and a diffusive behavior is observed.
On the other hand, energetic phonons ($\omega=$800, 1200, 1500 cm$^{-1}$) experience localization at different lengths depending on their energies as it is shown in Fig.~\ref{fig:regime}(b).

Different transport regimes coexist for phonons with different energies.
Altough high energy phonons experience localization, a pure localization regime or an exponential dependence of thermal conductance on CNT length is not observed.

Therefore heating of the semiconductor CNT device in electronic applications is dependent on the length of the device and the frequency distribution of generated phonons.
Low energy phonons will be discharged effectively while localization of energetic phonons may cause overheating of long devices.
Localization induced heating becomes more dramatic at high operating temperatures.

\begin{figure}[t]
	\begin{center}
	\includegraphics[width=0.9\textwidth]{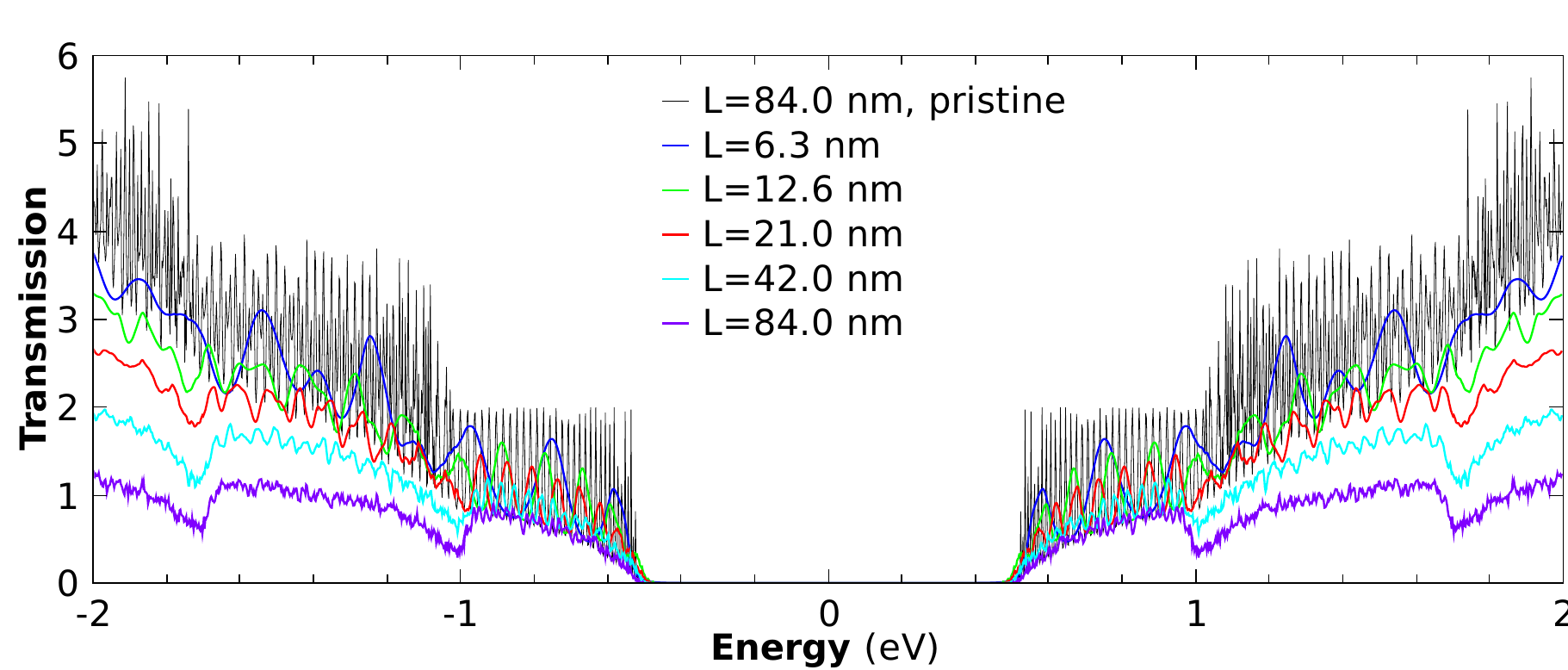}
	\caption{(Color online) Electron transport through CNT(10,0) with Anderson disorder. Transmission as a function of  energy is shown for different lengths.}
	\label{fig:spectra}
	\end{center}
\end{figure}

\begin{figure}[b]
	\begin{center}
	\includegraphics[width=0.8\textwidth]{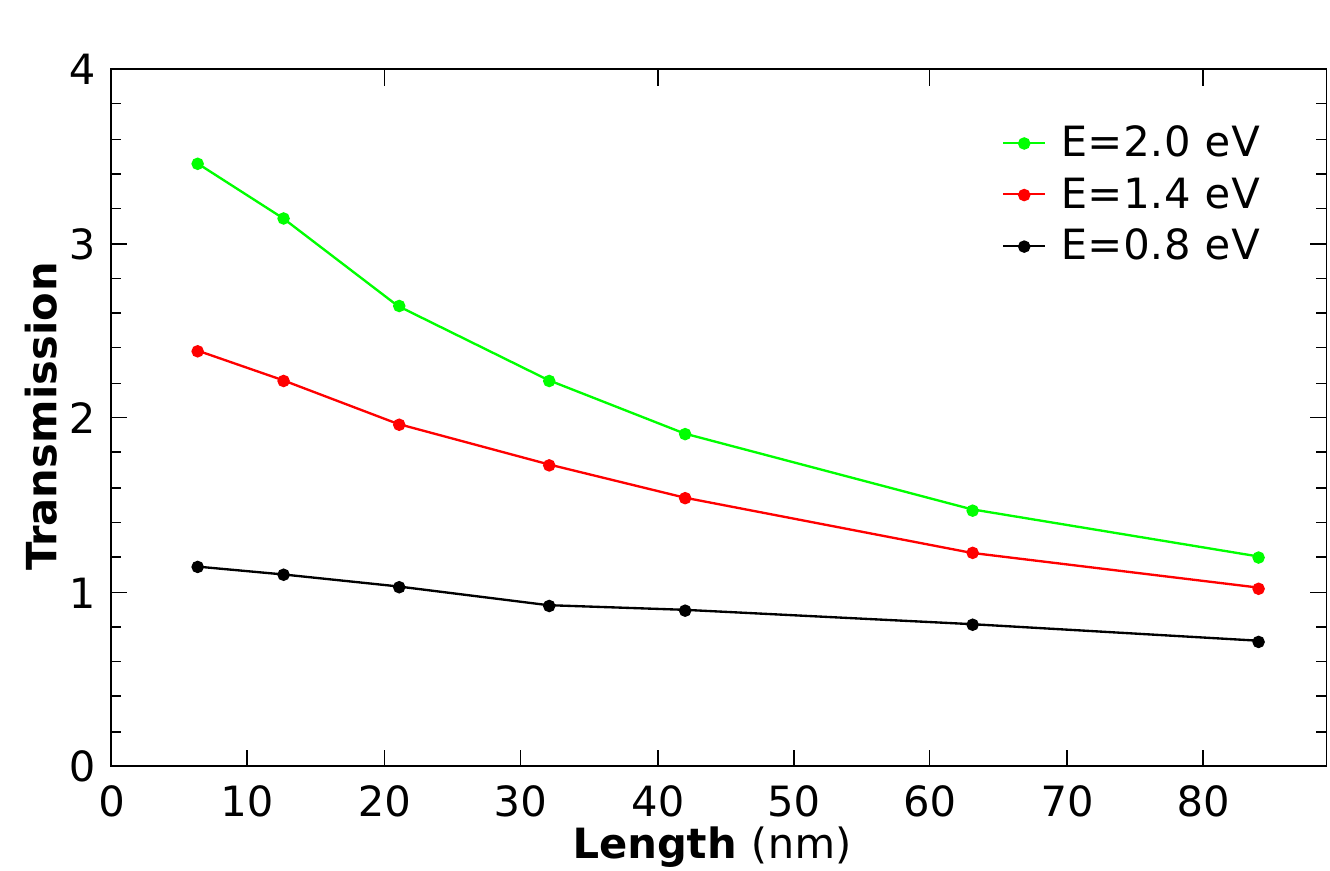}
	\caption{(Color online) Electron transport through CNT(10,0) with Anderson disorder. Transmission as a function of length  is shown for different energies.}
	\label{fig:T2}
	\end{center}
\end{figure}

The Anderson disorder for electron transport is introduced by random variation of the hopping matrix elements $t_{ij}$, which are sensitive to geometry fluctuations. We take the range of $t_{ij}$ fluctuations to be plus/minus 10\% of the average value (2.7 eV) corresponding to mass disorder. We average the curves over 50 configurations, the results are presented in Figs. \ref{fig:spectra} and \ref{fig:T2}. As the model for electrodes we take the wide-band metal edge electrodes coupled to the edge carbon atoms of nanotubes, so that strong quantum interference oscillations of the transmission are observed in the pristine case (Fig.\,\ref{fig:spectra}). Oppositely, transmission through disordered CNTs is more regular because of the self-averaging. The transmission as a function of length demonstrates exponential suppression at large lengths, which is a signature of localization in quasi one-dimensional system. In Fig.\,\ref{fig:T2} the length dependence at three different energies is shown, it is clear that scattering increases and the localization length decreases at larger energies, when large number of electron bands is involved.  

It is interesting to compare the energy dependence of disorder effect on electron and phonon transport. Phonon transport is not affected only at very low energies, so that at room temperature the thermal conductance is effectively suppressed. Electron transport at the energies within the first band of transverse quantization is not affected by moderately strong disorder of the considered type. It means that the conductance of doped nanotubes (which are semiconducting with large enough gap) or finite-voltage conductance due to injection of carriers into the conduction band will not be changed essentially by disorder, while the phonon conductance can be suppressed.   

\section{Conclusions}

In summary, we analyzed the effects of disorder on phonon thermal transport and electron charge transport in semiconductor CNTs. We show that different transport regimes for phonons of different energies coexist and their relative weight on thermal conductance depends on temperature. Long wavelength phonons are transmitted effectively, but short wavelength phonons are likely cause overheating and the effect of localization will be more pronounced at higher operating temperatures. On the other hand, electron transport is much less sensitive to disorder. One can conclude that disorder can be an effective mechanism of engineering the thermoelectric properties of carbon nanostructures, as was shown by two of us in Ref.\,\cite{Sevincli10prb} for graphene nanoribbons.

\begin{acknowledgement}
We acknowledge fruitful discussion with R. Gutierrez.
This work was supported by the European Union project ``Carbon nanotube devices at the quantum limit'' (CARDEQ) under contract No. IST-021285-2, by the Deutsche Forschungsgemeinschaft within the priority program ``Nanostructured Thermoelectrics'' (Grant No. SPP 1386) and from the German Excellence Initiative via the Cluster of Excellence 1056 ``Center for Advancing Electronics Dresden'' (cfAED). Computing time provided by the ZIH at the Dresden University of Technology is acknowledged.
\end{acknowledgement}


\begin{thebibliography}{10}
\providecommand{\url}[1]{{#1}}
\providecommand{\urlprefix}{URL }
\expandafter\ifx\csname urlstyle\endcsname\relax
  \providecommand{\doi}[1]{DOI~\discretionary{}{}{}#1}\else
  \providecommand{\doi}{DOI~\discretionary{}{}{}\begingroup
  \urlstyle{rm}\Url}\fi

\bibitem{Latil04prl}
Latil, S., Roche, S., Mayou, D., Charlier, J.-C.:
Mesoscopic Transport in Chemically Doped Carbon Nanotubes
\newblock Phys. Rev. Lett. \textbf{92}, 256805
\newblock  (2004)

\bibitem{Roche05prb}
Roche, S., Jiang, J., Triozon, F., Saito, R.:
Conductance and coherence lengths in disordered carbon nanotubes: Role of lattice defects and phonon vibrations
\newblock Phys. Rev. B \textbf{72}, 113410
\newblock  (2005)

\bibitem{Charlier07rmp}
Charlier, J.-C., Blase, X., Roche, S.:
Electronic and transport properties of nanotubes
\newblock Rev. Mod. Phys. \textbf{79}, 677
\newblock  (2007)

\bibitem{Lherbier08prl}
Lherbier, A., Biel, B., Niquet, Y.-M., Roche, S.:
Transport Length Scales in Disordered Graphene-Based Materials: Strong Localization Regimes and Dimensionality Effects
\newblock Phys. Rev. Lett. \textbf{100}, 036803
\newblock  (2008)

\bibitem{Dresselhaus08jvacscitechb}
Dresselhaus, M.S., Dresselhaus, G., Hofmann, M.: Other one-dimensional systems
  and thermal properties.
\newblock J. Vac. Sci. Technol. B \textbf{26}, 1613
\newblock  (2008)

\bibitem{Hochbaum08nature}
Hochbaum, A.I., Chen, R., Delgado, R.D., Liang, W., Garnett, E.C., Najarian,
  M., Majumdar, A., Yang, P.: Enhanced thermoelectric performance of rough
  silicon nanowires.
\newblock Nature \textbf{451}, 163
\newblock  (2008)

\bibitem{Boukai08nature}
Boukai, A.I., Bunimovich, Y., Tahir-Kheli, J., Yu, J.K., Goddard, W.A., Heath,
  J.R.: Silicon nanowires as efficient thermoelectric materials.
\newblock Nature \textbf{451}, 168
\newblock  (2008)

\bibitem{Markussen09prb}
Markussen, T., Jauho, A.P., Brandbyge, M.: Electron and phonon transport in
  silicon nanowires: Atomistic approach to thermoelectric properties.
\newblock Phys. Rev. B \textbf{79}, 035415
\newblock  (2009)

\bibitem{Chang07apl}
Chang, C.W., Okawa, D., Garcia, H., Yuzvinsky, T.D., Majumdar, A., Zettl, A.:
  Tunable thermal links.
\newblock Appl. Phys. Lett. \textbf{90}, 193114
\newblock  (2007)

\bibitem{Savic08prl}
Savi\ifmmode~\acute{c}\else \'{c}\fi{}, I., Mingo, N., Stewart, D.A.: Phonon
  transport in isotope-disordered carbon and boron-nitride nanotubes: Is
  localization observable?
\newblock Phys. Rev. Lett. \textbf{101}, 165502
\newblock  (2008)

\bibitem{Stewart09nanolett}
Stewart, D.A., Savić, I., Mingo, N.: First-principles calculation of the
  isotope effect on boron nitride nanotube thermal conductivity.
\newblock Nano Letters \textbf{9}, 81
\newblock  (2009)

\bibitem{Chang06prb}
Chang, C.W., Fennimore, A.M., Afanasiev, A., Okawa, D., Ikuno, T., Garcia, H.,
  Li, D., Majumdar, A., Zettl, A.: Isotope effect on the thermal conductivity
  of boron nitride nanotubes.
\newblock Phys. Rev. Lett. \textbf{97}, 085901
\newblock  (2006)

\bibitem{Wallace47prb}
Wallace, P.R.: The band theory of graphite.
\newblock Phys. Rev. \textbf{71}, 622
\newblock  (1947)

\bibitem{LopezSancho84jpf}
{Lopez~Sancho}, M.P., {Lopez~Sancho}, J.M., Rubio, J.: Quick iterative scheme
  for the calculation of transfer matrices: application to {Mo} (100).
\newblock J. Phys. F: Met. Phys. \textbf{14}, 1205
\newblock  (1985)

\bibitem{LopezSancho85jpf}
{Lopez~Sancho}, M.P., {Lopez~Sancho}, J.M., Rubio, J.: Highly convergent
  schemes for the calculation of bulk and surface green functions.
\newblock J. Phys. F: Met. Phys. \textbf{15}, 851
\newblock  (1985)

\bibitem{Datta95book}
Datta, S.: Electronic Transport in Mesoscopic Systems.
\newblock Cambridge University Press, Cambridge
\newblock  (1995)

\bibitem{Cuniberti05book}
Cuniberti, G., Fagas, G., {Richter (Eds.)}, K.: Introducing Molecular
  Electronics, \emph{Lecture Notes in Physics}, vol. 680.
\newblock Springer, Berlin
\newblock  (2005)

\bibitem{Ryndyk09inbook}
Ryndyk, D.A., Guti\'{e}rrez, R., Song, B., Cuniberti, G.: Energy Flow Dynamics
  in Biomaterial Systems, chap. Green function techniques in the treatment of
  quantum transport at the molecular scale, p. 213.
\newblock Springer, Berlin
\newblock  (2009)

\bibitem{Cuevas10book}
Cuevas, J.C., Scheer, E.: Molecular electronics: An Introduction to Theory and
  Experiment.
\newblock World Scientific
\newblock  (2010)

\bibitem{Gunlycke07prb}
Gunlycke, D., Lawler, H.M., White, C.T.: Room-temperature ballistic transport
  in narrow graphene strips.
\newblock Phys. Rev. B \textbf{75}, 085418
\newblock  (2007)

\bibitem{Saito88book}
Saito, R., Dresselhaus, G., Dresselhaus, M.S.: Physical Properties of Carbon
  Nanotubes.
\newblock Imperial College Press, London
\newblock  (1988)

\bibitem{Zimmermann08prb}
Zimmermann, J., Pavone, P., Cuniberti, G.: Vibrational modes and
  low-temperature thermal properties of graphene and carbon nanotubes: Minimal
  force-constant model.
\newblock Phys. Rev. B \textbf{78}, 045410
\newblock  (2008)

\bibitem{Sevincli0905.3815}
Sevin\c{c}li, H., Cuniberti, G.: Heat conduction in disordered semiconductor
  carbon nanotubes.
\newblock arXiv:0905.3815 (unpublished).
\newblock  (2009)

\bibitem{Yamada08prb}
Yamada, M., Yamakita, Y., Ohno, K.: Phonon dispersions of hydrogenated and
  dehydrogenated carbon nanoribbons.
\newblock Phys. Rev. B \textbf{77}, 054302
\newblock  (2008)

\bibitem{Angelescu98sm}
Angelescu, D., Cross, M., Roukes, M.: Heat transport in mesoscopic systems.
\newblock Superlattices and Microstructures \textbf{23}, 673
\newblock  (1998)

\bibitem{Rego98prl}
Rego, L.G.C., Kirczenow, G.: Quantized thermal conductance of dielectric
  quantum wires.
\newblock Phys. Rev. Lett. \textbf{81}, 232
\newblock  (1998)

\bibitem{Ozpineci01prb}
Ozpineci, A., Ciraci, S.: Quantum effects of thermal conductance through atomic
  chains.
\newblock Phys. Rev. B \textbf{63}, 125415
\newblock  (2001)

\bibitem{Mingo03prb}
Mingo, N., Yang, L.: Phonon transport in nanowires coated with an amorphous
  material: An atomistic {G}reen's function approach.
\newblock Phys. Rev. B \textbf{68}, 245406
\newblock  (2003)

\bibitem{Segal03jcp}
Segal, D., Nitzan, A., H\"{a}nggi, P.: Thermal conductance through molecular
  wires.
\newblock The Journal of Chemical Physics \textbf{119}, 6840
\newblock  (2003)

\bibitem{Mingo06prb}
Mingo, N.: Anharmonic phonon flow through molecular-sized junctions.
\newblock Phys. Rev. B \textbf{74}, 125402
\newblock  (2006)

\bibitem{Wang06prb}
Wang, J.S., Wang, J., Zeng, N.: Nonequilibrium green's function approach to
  mesoscopic thermal transport.
\newblock Phys. Rev. B \textbf{74}, 033408
\newblock  (2006)

\bibitem{Yamamoto06prb}
Yamamoto, T., Watanabe, K.: Nonequilibrium green's function approach to phonon
  transport in defective carbon nanotubes.
\newblock Phys. Rev. Lett. \textbf{96}, 255503
\newblock  (2006)

\bibitem{Galperin07prb2}
Galperin, M., Nitzan, A., Ratner, M.A.: Heat conduction in molecular transport
  junctions.
\newblock Phys. Rev. B \textbf{75}, 155312
\newblock  (2007)

\bibitem{Wang08epjb}
Wang, J.S., Wang, J., L\"{u}, J.T.: Quantum thermal transport in
  nanostructures.
\newblock The European Physical Journal B \textbf{62}, 381
\newblock  (2008)

\bibitem{Dubi11rmp}
Dubi, Y., Di~Ventra, M.: Heat flow and thermoelectricity in atomic and
  molecular junctions.
\newblock Rev. Mod. Phys. \textbf{83}, 131
\newblock  (2011)

\bibitem{Nikolic12jcel}
Nikoli\'{c}, B.K., Saha, K.K., Markussen, T., Thygesen, K.S.: First-principles
  quantum transport modeling of thermoelectricity in single-molecule
  nanojunctions with graphene nanoribbon electrodes.
\newblock Journal of Computational Electronics \textbf{11}, 78
\newblock  (2012)

\bibitem{Meir92prl}
Meir, Y., Wingreen, N.S.: Landauer formula for the current through an
  interacting electron region.
\newblock Phys. Rev. Lett. \textbf{68}, 2512
\newblock  (1992)

\bibitem{Sevincli10prb}
Sevin\ifmmode~\mbox{\c{c}}\else \c{c}\fi{}li, H., Cuniberti, G.: Enhanced
  thermoelectric figure of merit in edge-disordered zigzag graphene
  nanoribbons.
\newblock Phys. Rev. B \textbf{81}, 113401
\newblock  (2010)

\end{thebibliography}
\end{document}